\documentclass[slac_one]{revtex4}
\usepackage[pdftex]{graphicx}
\usepackage{fancyhdr}
\def\ra    {\rightarrow}

\def\Du    {\ensuremath{D^0}}
\def\bDu   {\ensuremath{\bar{D}^0}}
\def\Dcp   {\ensuremath{D_{CP}^0}}

\def\Bd    {\ensuremath{B^0}}
\def\Bu    {\ensuremath{B^+}}
\def\bBu   {\ensuremath{B^-}}

\def\BuJpsiK   {\ensuremath{\Bu \ra J/\psi\,K^+}}
\def\BuDK  {\ensuremath{\Bu\ra\bDu\,K^+}}
\def\bBuDK {\ensuremath{\bBu\ra\Du\,K^-}}
\def\BuDcpK  {\ensuremath{\Bu\ra D_{CP}^0\,K^+}}
\def\bBuDcpK {\ensuremath{\bBu\ra D_{CP}^0\,K^-}}
\def\BuDPi  {\ensuremath{\Bu\ra\bDu\,\pi^+}}
\def\bBuDPi {\ensuremath{\bBu\ra\Du\,\pi^-}}
\def\BuDcpPi  {\ensuremath{\Bu\ra D_{CP}^0\,\pi^+}}
\def\bBuDcpPi {\ensuremath{\bBu\ra D_{CP}^0\,\pi^-}}

\def\mevc  {\ifmmode {\rm MeV}/c \else MeV$/c$\fi}
\def\mevcc {\ifmmode {\rm MeV}/c^2 \else MeV$/c^2$\fi}
\def\gevc  {\ifmmode {\rm GeV}/c \else GeV$/c$\fi}
\def\gevcc {\ifmmode {\rm GeV}/c^2 \else GeV$/c^2$\fi}
\def\lxy   {\ifmmode L_{\rm xy} \else $L_{\rm xy}$\fi}
\def\dedx  {\ensuremath{dE/dx}}
\def\fb    {\ifmmode {\rm fb}^{-1} \else fb$^{-1}$\fi}
\def\Rcp   {\ensuremath{R_{CP+}}}
\def\Acp   {\ensuremath{A_{CP+}}}

\def\Babar {\mbox{\sl B\hspace{-0.4em} {\small\sl A}\hspace{-0.37em} \sl B\hspace{-0.4em} {\small\sl A\hspace{-0.02em}R}}}

\begin{document}

\title{\boldmath Measurement of $\textit{CP}$ Observables in $\bBuDK$ Decays at CDF}

\author{Karen~R.~Gibson on behalf of the CDF Collaboration}
\affiliation{University of Pittsburgh, Pittsburgh, Pennsylvania 15260}


\begin{abstract}
This paper describes the determination of the direct $\textit{CP}$
observables $\Rcp$ and $\Acp$, related to the CKM angle $\gamma$,
using $\bBuDK$ decays in 1.0~$\fb$ of data collected with the CDF\,II
detector.  This is the first measurement of these
$\textit{CP}$~observables made at a hadron collider in this decay
mode.  The results presented are consistent with the measurements of
these quantities made by the Belle and $\Babar$~collaborations.
\end{abstract}

\maketitle

\thispagestyle{fancy}


\section{INTRODUCTION}

Over the past twenty years, there have been many efforts to determine
the length of the sides and the angles of the well-known
Cabibbo-Kobayashi-Maskawa (CKM) unitarity triangle for light $B$
mesons~\cite{Ref:PDG}, which is constructed from the first and third
columns of the CKM quark mixing matrix~\cite{Ref:Cabibbo}.  At
present, good knowledge of the CKM angle $\beta$ has been established
by the Belle and $\Babar$ collaborations~\cite{Ref:HFAG}.  However,
comparable knowledge of the other CKM angles $\gamma$ and $\alpha$ is
lacking, due to limited statistics in the decay channels sensitive to
these angles, such as $\bBuDK$ in the case of $\gamma$ and
$\Bd\ra\pi^+\pi^-$ in the case of $\alpha$.

This paper concerns the determination of $\textit{CP}$~observables in
the decay $\bBuDK$~\cite{Ref:CC}, which provide sensitivity to the
angle $\gamma\equiv\arg((-V_{ud}V^*_{ub})/(V_{cd}V^*_{cb}))$, using
the method developed by Michael Gronau, David London and Daniel Wyler
(GLW) in the early 1990s~\cite{Ref:GLW}.  These $\textit{CP}$
observables are constructed from the measurement of relative branching
ratios and from $\textit{CP}$~asymmetries.  Previously, measurements
of these quantities, and of $\gamma$ by extension, have only been made
at the $B$~factories.  Since these measurements have been limited by
statistical accuracy, the experiments at the Tevatron have an
opportunity to contribute significantly to the determination of these
quantities, particularly when considering the large amount of
$B$~hadron data that remains to be collected and analyzed before the
end of Run\,II.  This paper presents the first measurement of the
$\textit{CP}$~observables constructed in $\bBuDK$ decays at a hadron
collider, using 1.0~$\fb$ of data that was collected between February
2002 and February 2006.

\section{GLW \boldmath{$\textit{CP}$} OBSERVABLES}

The GLW method of determining $\gamma$ was developed as a way to
reduce the theoretical error on the measurement of $\gamma$ in
$\bBuDK$ decays by using $\textit{CP}$-even or odd decays of the
$\Du$.  The asymmetry between the $\bBu$ and $\Bu$ amplitudes of
decays to a $\Du$ $\textit{CP}$~eigenstate can be related to the
amplitudes of decays to a flavor-specific $\Du$ decay and the angle
$\gamma$
\begin{equation}
|A(\bBuDcpK)|^2 - |A(\BuDcpK)|^2 = 2|A(\bBu\ra\bDu K^-)||A(\bBuDK)|\sin\delta\sin\gamma,
\label{eqn:GLW}
\end{equation}
where $\delta$ is the relative final-state interaction phase of the
amplitudes to flavor-specific $\Du$ decays.  From the right-hand side
of Eq.~(\ref{eqn:GLW}), it can be seen that there are two possible
solutions for $\gamma$.

In the present measurement, only $\textit{CP}$-even decays of $\Du$
are considered, $\Du\ra K^+K^-$ and
$\Du\ra\pi^+\pi^-$~\cite{Ref:CPodd}.  The rates of these decays are
normalized to the flavor specific decay $\Du\ra K^-\pi^+$ through
following experimental observables
\begin{eqnarray}
R  &=&\frac{BR(\bBuDK)+BR(\BuDK)}{BR(\bBuDPi)+BR(\BuDPi)}, \label{eqn:GLW1}\\
R_+&=&\frac{BR(\bBuDcpK)+BR(\BuDcpK)}{BR(\bBuDcpPi)+BR(\BuDcpPi)}.
\label{eqn:GLW2}
\end{eqnarray}
We can then relate these observables to the $\textit{CP}$~observables 
\begin{eqnarray}
\Rcp&=&\frac{R_+}{R} = 1 + r^2 + 2r\cos\delta\cos\gamma, \\
\Acp&=&\frac{BR(\bBuDcpK)-BR(\BuDcpK)}{BR(\bBuDcpK)+BR(\BuDcpK)}\nonumber\\
    &=&2r\sin\delta\sin\gamma/\Rcp.
\end{eqnarray}
where $r = |A(\bBu\ra\bDu K^-)|/|A(\bBuDK)|$.  

\section{MEASUREMENT OF \boldmath{$\textit{CP}$} OBSERVABLES AT CDF}

The CDF\,II detector is described in detail
elsewhere~\cite{Ref:fqPRD}.  The data used in the present measurement
is collected with a track-based trigger that requires two displaced
tracks to satisfy individual and combined minimum transverse momenta
thresholds~\cite{Ref:TTT}.  The relative branching ratios in
Eqs.~(\ref{eqn:GLW1}) and~(\ref{eqn:GLW2}) are measured in a
simultaneous maximum likelihood fit, which uses both kinematic
information, such as the invariant mass of the $\bBu$ system, which is
reconstructed as $\Du\pi^-$, and daughter particle momenta, and the
specific ionization of the daughter track from the $\bBu$ in order to
separate kaons and pions.

To choose the signal selection requirements, the sensitivity to the
primary quantity of interest, $\Acp$, is optimized using
pseudo-experiments.  A likelihood fit is performed in the invariant
mass window $m(\Du\pi^-)\in[5.17, 5.60]~\gevcc$.  This requirement
reduces backgrounds from decays other than $\bBu\ra D^{(*)0}\pi^-$.
Additionally, events which fall within a $\pm 2\sigma$ window around
the $\BuJpsiK$ are removed from the sample to reduce contamination in
the $\bBu\ra\Du[\ra\pi^-\pi^+]K^-$.  However, it is worthwhile to note
that background events from $B^-\ra K^-K^+K^-$ cannot be eliminated
and must be included in the likelihood fit.

\begin{figure}[htbp]
\centerline{
\makebox{\includegraphics[width=0.45\textwidth]{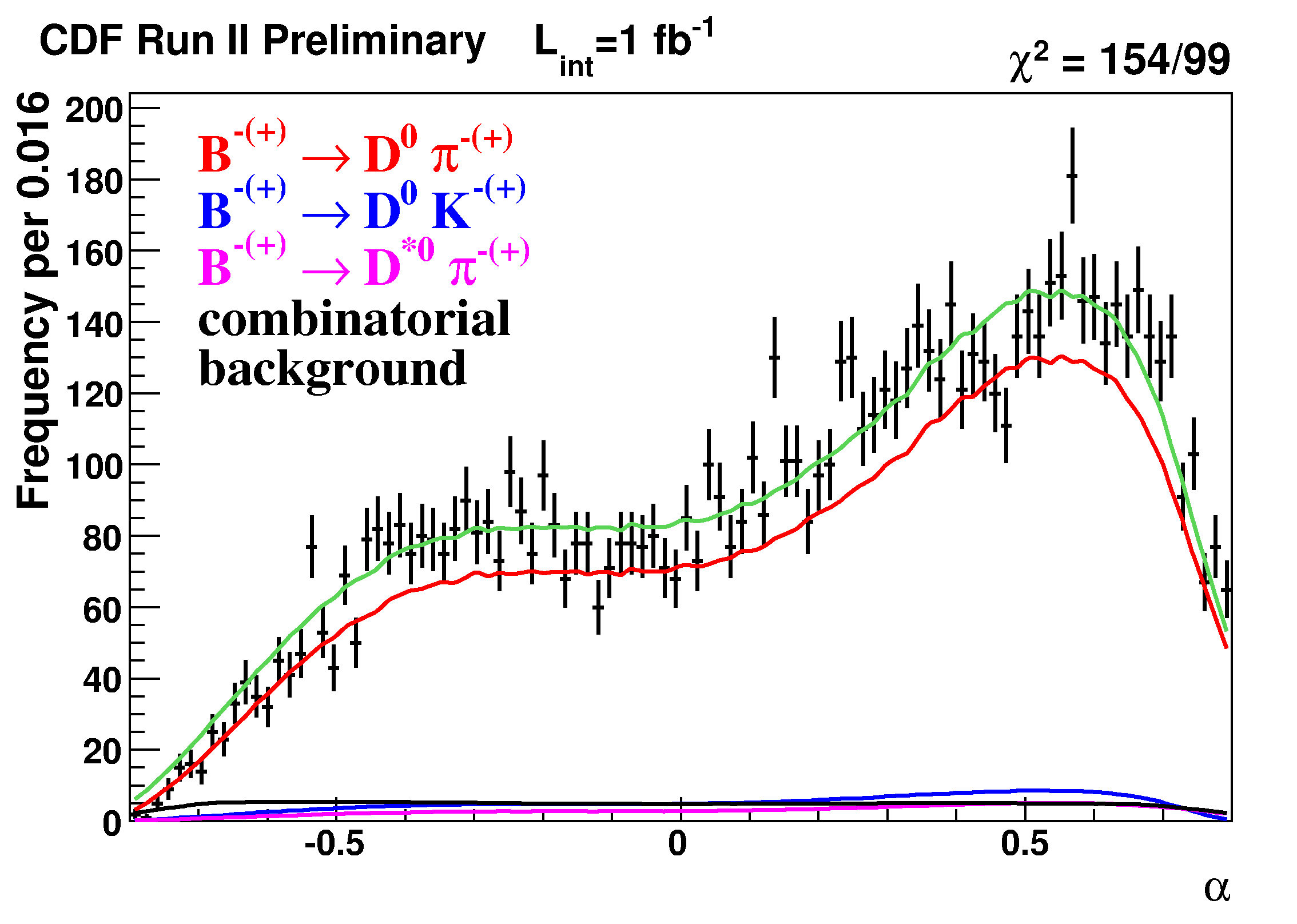}}
\makebox{\includegraphics[width=0.45\textwidth]{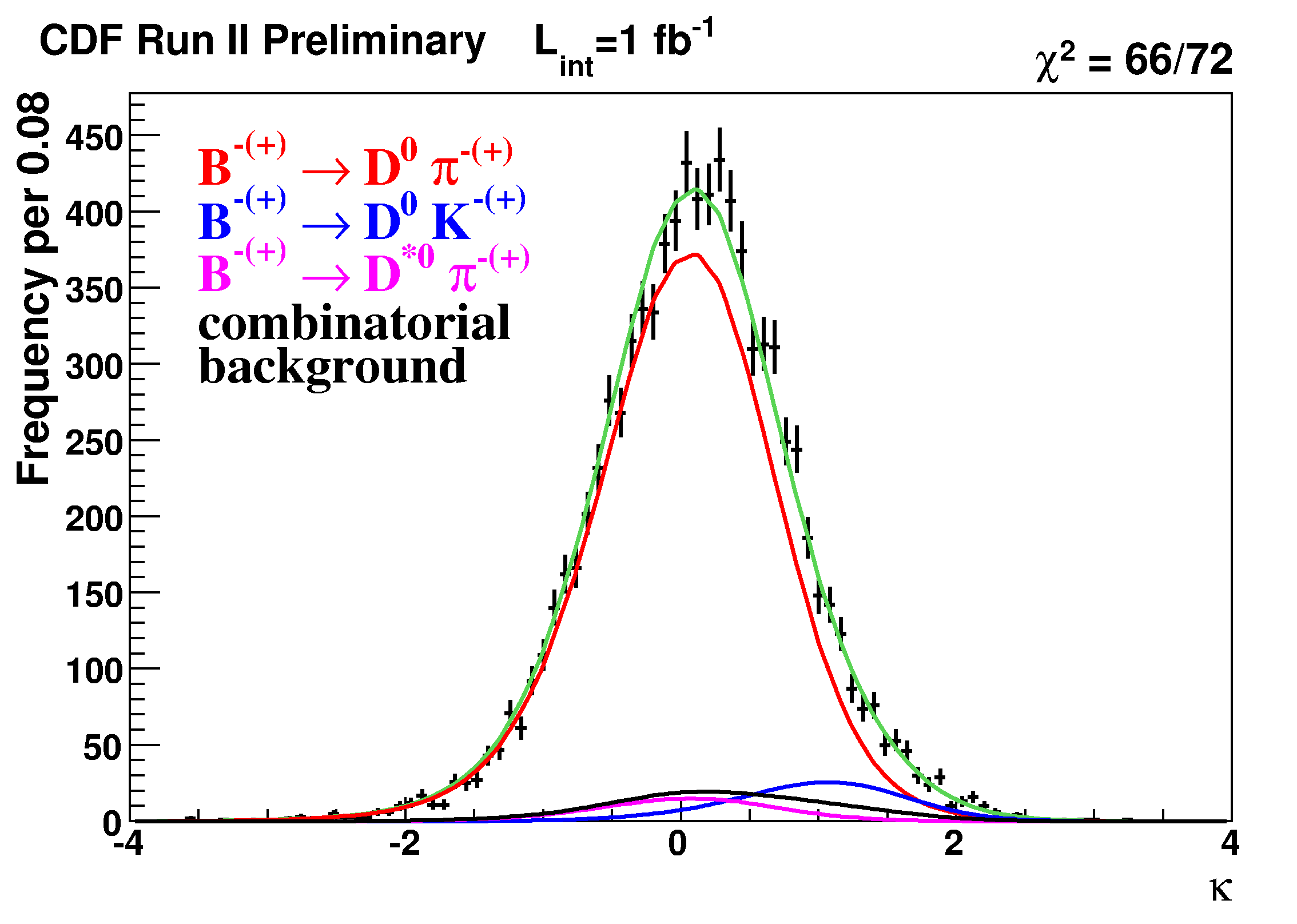}}
}
\caption{ Fit projections of $\alpha$ (left) and $\kappa$ (right) for
$\Bu\ra[K^-\pi^+]\pi^+$ decays. The green line in both plots indicates
the overall fit.}
\label{fig:proj}
\end{figure}

Six identical likelihoods are constructed, one set each for $\Bu$ and
$\bBu$ and three in each set corresponding to the three $\Du$ decays
considered, $\Dcp\ra K^-K^+$, $\Dcp\ra \pi^-\pi^+$, and $\Du\ra
K^-\pi^+$.  $-2\ln{\cal L}$ is simultaneously minimized over all six
likelihoods ${\cal L}_i$, which are written in terms of ``signal''
probability distribution functions (PDFs) $F_{\Du K}$, $F_{\Du\pi}$,
and $F_{D^{*0}\pi}$ and a background PDF $F_{bg}$ as
\begin{eqnarray}
{\cal L}_i&=&\prod_{k}^{N_{ev}}
((1-f_{bg})((1-f_{\Du\pi}-f_{D^{*0}\pi})F_{\Du K}(\alpha,p_{tot},m(\Du\pi),\kappa) \nonumber\\
          &&+f_{\Du\pi}F_{\Du\pi}(\alpha,p_{tot},m(\Du\pi),\kappa)
            +f_{D^{*0}\pi}F_{D^{*0}\pi}(\alpha,p_{tot},m(\Du\pi),\kappa)) \nonumber\\
          &&+f_{bg}F_{bg}(\alpha,p_{tot},m(\Du\pi),\kappa)),
\label{eqn:likelihood}
\end{eqnarray}
where the inputs to the fit are the invariant mass $m(D^0\pi^-)$, the
momentum of the $\Du$ plus track $p_{tot} = p_{\pi/K}+p_{D^0}$, the
momentum imbalance $\alpha = 1-p_{\pi/K}/p_{\Du}~\mbox{if}~p_{\pi/K} <
p_{\Du}$ and $\alpha=-(1-p_{\Du}/p_{\pi/K})~\mbox{if}~p_{\pi/K} \ge
p_{\Du}$, and the track flavor identification variable $\kappa =
((\dedx)_{meas} - (\dedx)_{exp}(\pi))/((\dedx)_{exp}(K) -
(\dedx)_{exp}(\pi))$.  The fit projections in $\alpha$ and $\kappa$
are shown in Fig.~\ref{fig:proj} for $\Du\ra K^-\pi^+$ decays.  The
background PDF in Eq.~(\ref{eqn:likelihood}) is expanded for the
decays of $\Dcp\ra K^-K^+$ in order to separate the combinatoric
background from the background due to $\bBu\ra K^-K^+K^-$ decays.
Simulated events are used to model correlations between the variables
$\alpha$ and $p_{tot}$ in $F_{\Du K}$, $F_{\Du\pi}$, and
$F_{D^{*0}\pi}$, while the upper $\bBu$ mass sideband is used to model
the correlation in the background.

\begin{figure}[htbp]
\centerline{
\makebox{\includegraphics[width=0.45\textwidth]{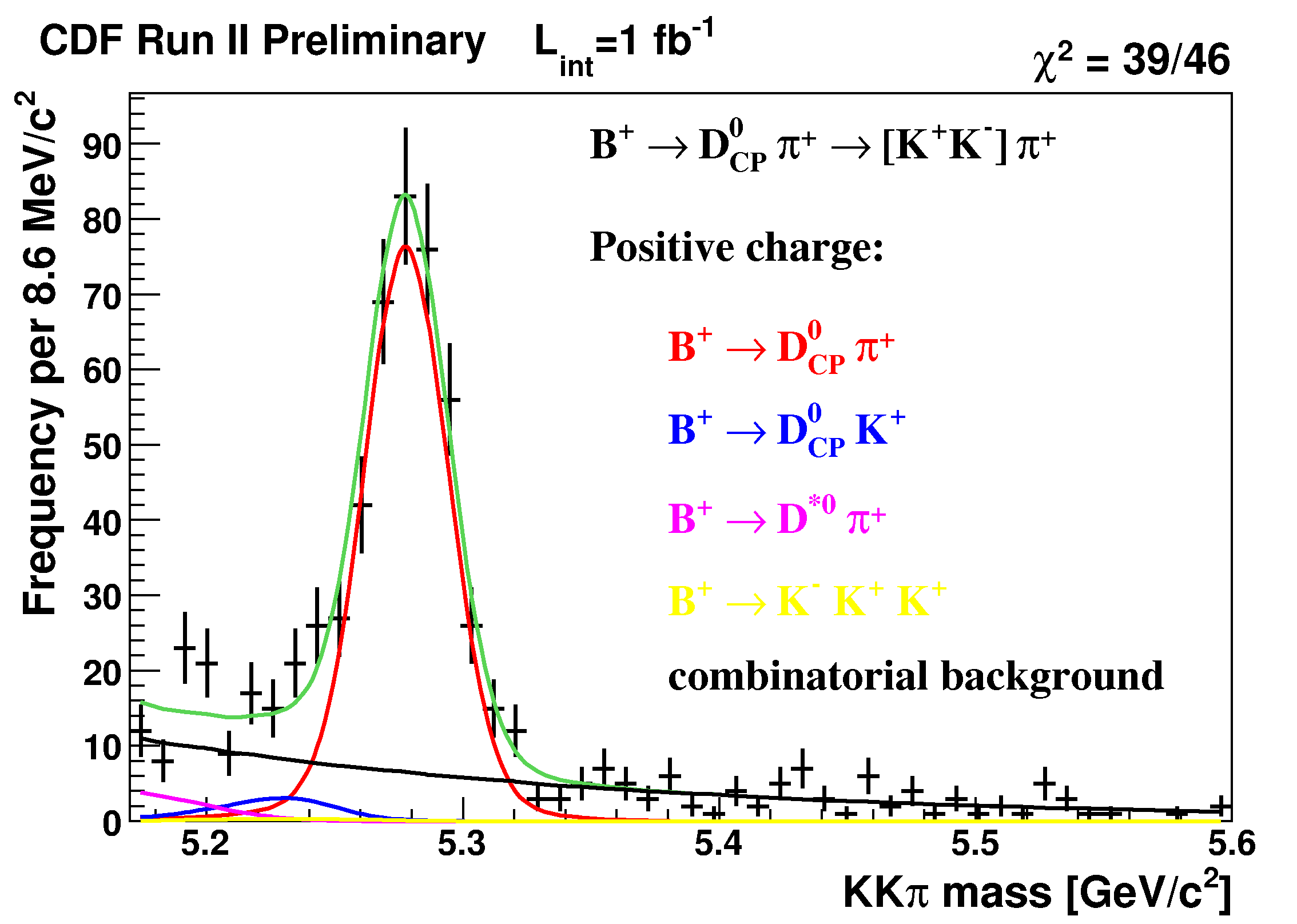}}
\makebox{\includegraphics[width=0.45\textwidth]{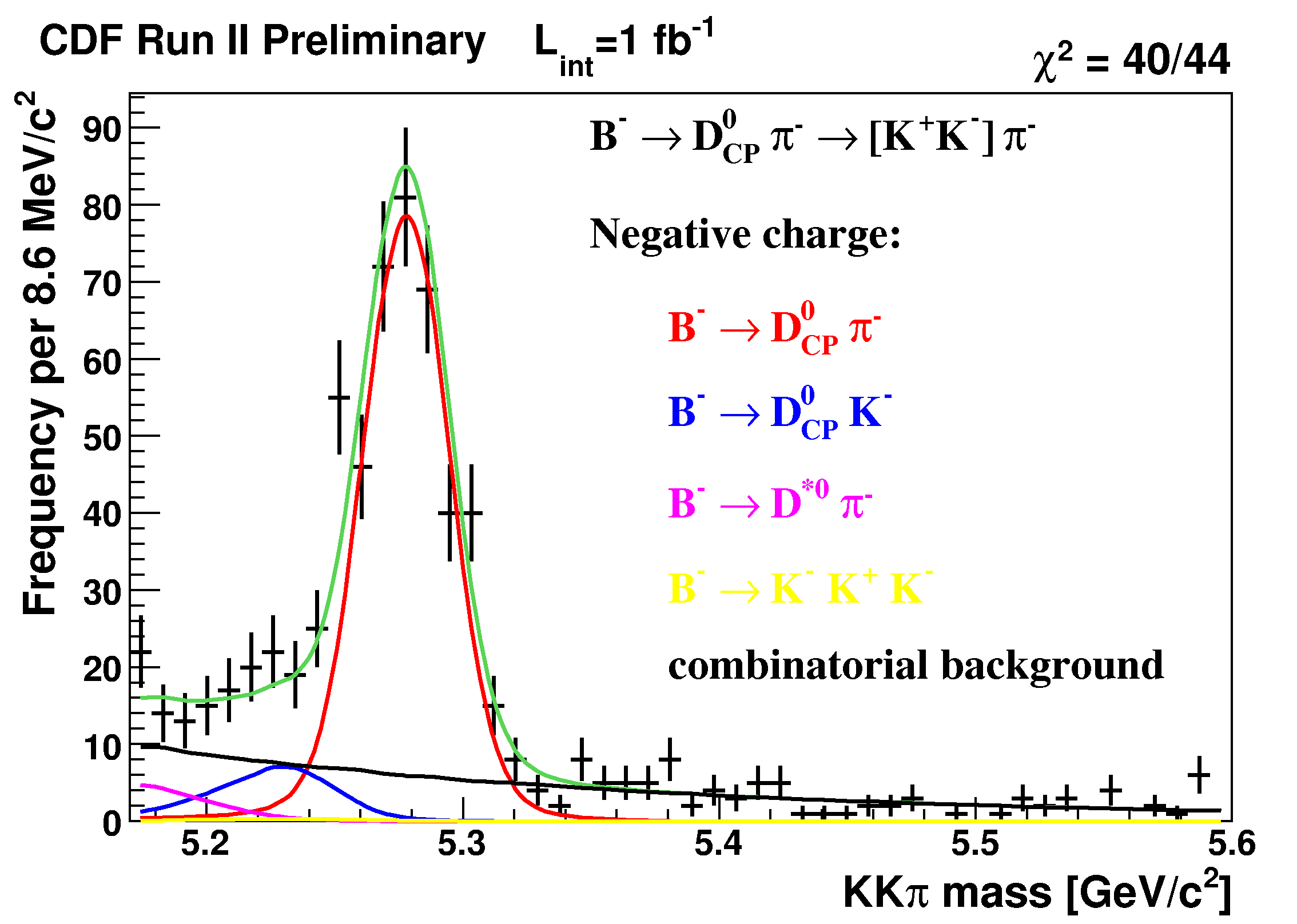}}
}
\caption{ Mass fit projections for $\Bu\ra[K^-K^+]\pi^+$ (left) and
$\bBu\ra[K^-K^+]\pi^-$ (right). The green line in both plots indicates
the overall fit.}
\label{fig:yields}
\end{figure}

From the likelihood fit, we obtain raw yields for the six decays, two
of which are shown in Fig.~\ref{fig:yields}.  The ratios $R$ and
$\Acp$ obtained from these yields are corrected using Monte Carlo
simulation to account for the relative efficiencies between the decay
modes.  Small biases observed in pseudo-experiments, on the level of
1-2\%, are corrected for as well.

\section{SYSTEMATIC UNCERTAINTIES}

A number of sources of systematic uncertainty are evaluated for the
ratios $R$, $\Rcp$, and $\Acp$.  The largest sources of systematic
uncertainty are due to particle flavor identification ($\dedx$) and
the correlation between $\alpha$ and $p_{tot}$ in the combinatoric
background.  The second of these systematic uncertainties is limited
by the statistics of the invariant mass sidebands, so its knowledge
will improve with increased data.  The description of the invariant
mass of the $D^{*0}\pi^-$ is also a significant source of systematic
uncertainty in the measurement of $R$.  Other sources of systematic
uncertainties considered are the uncertainty on the value of the
$\bBu$ mass~\cite{Ref:Bmass}, the mass model of the combinatoric
background, the knowledge of correlations between $\alpha$ and
$p_{tot}$ in the $\Du\pi^-$, $D^{*0}\pi^-$, and $\Du K^-$ PDFs, the
small bias observed in pseudo-experiments, and the statistics of the
simulated sample used to correct relative efficiencies.  The
systematic uncertainties assigned are listed in Tab.~\ref{tab:sys}.
\begin{table}[t]
\begin{center}
\begin{tabular}{|l|ccc|}
\hline
Source of systematic uncertainty              & $R$       & $\Rcp$       & $\Acp$ \\
\hline
$\dedx$                                       & 0.0028    & 0.056        & 0.030  \\
Input $\bBu$ mass                             & 0.0002    & 0.004        & 0.002  \\
$D^{*0}\pi^-$ mass model                      & 0.0028    & 0.025        & 0.006  \\
Combinatoric bg. mass model                   & 0.0002    & 0.020        & 0.001  \\
($\alpha$, $p_tot$) model of combinatoric bg. & 0.0002    & 0.100        & 0.020  \\
($\alpha$, $p_tot$) model of $\Du\pi^-$       & 0.0001    & 0.002        & 0.001  \\
($\alpha$, $p_tot$) model of $D^{*0}\pi^-$    & 0.0007    & 0.004        & 0.002  \\
($\alpha$, $p_tot$) model of $\Du K^-$        & 0.0006    & 0.002        & 0.004  \\
Pseudo-experiment bias error                  & 0.0001    & 0.005        & 0.003  \\
Monte Carlo simulation statistics             & 0.002     & -            & -      \\
\hline
{\textbf Total}                            & {\bf 0.0045} & {\bf 0.12} & {\bf 0.04} \\
\hline
\end{tabular}
\caption{\label{tab:sys}
Systematic uncertainties assigned to the measurement.}
\end{center}
\end{table}

\section{RESULTS}

The final results for $R$, $\Rcp$ and $\Acp$, corrected for relative
efficiencies and including systematic uncertainties, are
\begin{eqnarray*}
R&=&0.0745 \pm 0.0043~\mbox{(stat)} \pm  0.0043~\mbox{(syst)}, \\
\Rcp&=&1.30 \pm 0.24~\mbox{(stat)} \pm  0.12~\mbox{(syst)}, \\
\Acp&=&0.39 \pm 0.17~\mbox{(stat)} \pm  0.04~\mbox{(syst)}. \\
\label{eqn:results}
\end{eqnarray*}
\begin{figure}[tbp]
\centerline{
\makebox{\includegraphics[width=0.45\textwidth]{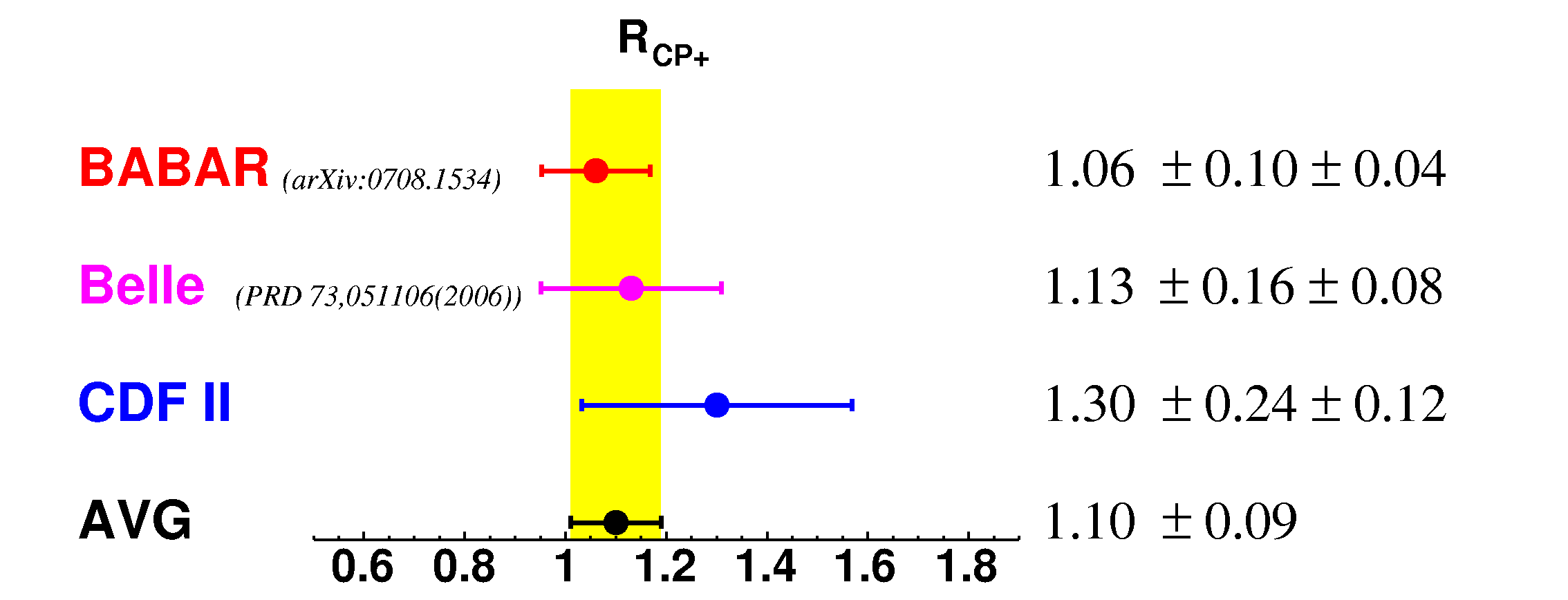}}
\makebox{\includegraphics[width=0.45\textwidth]{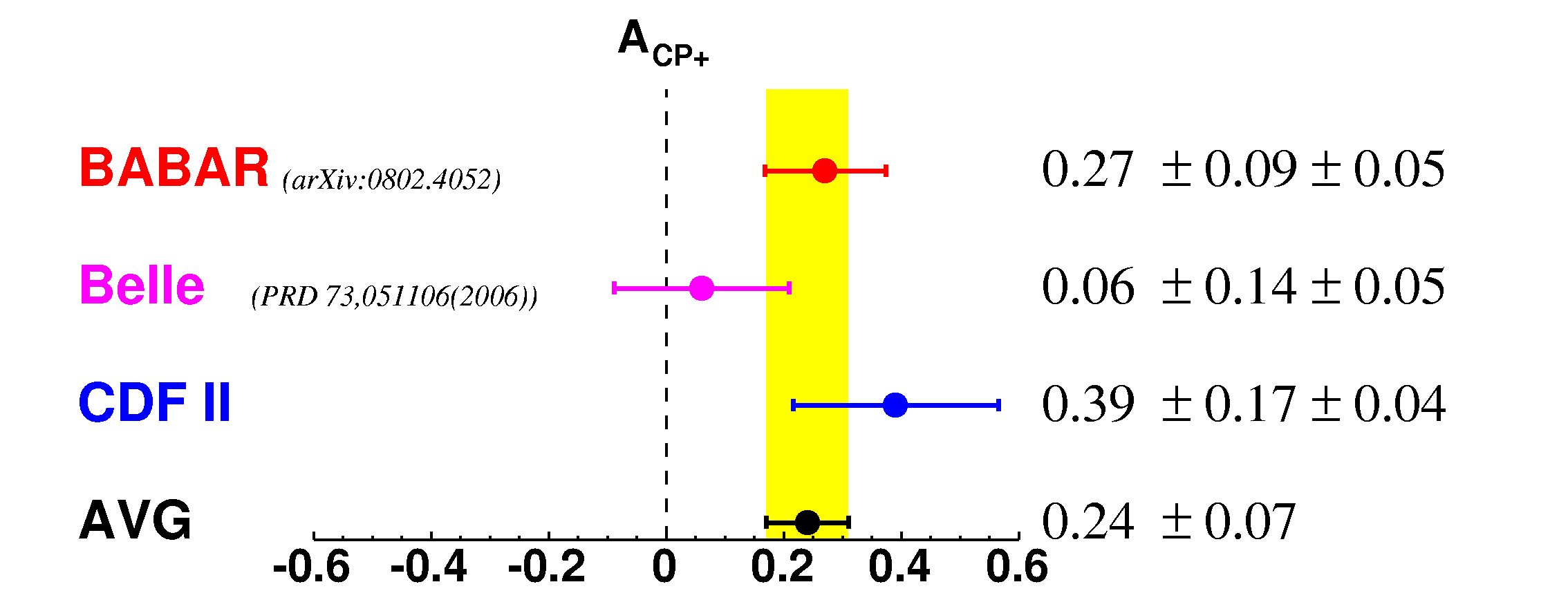}}
}
\caption{ 
Comparison of $\Rcp$ (left) and $\Acp$ (right) results with those from
Belle and $\Babar$.}
\label{fig:results}
\end{figure}
Additionally, $\Rcp$ and $\Acp$ are compared with the results from
Belle and $\Babar$ in Fig.~\ref{fig:results}.  This measurement
represents a significant achievement for a hadron collider experiment,
as it is the first time these $\textit{CP}$ observables have been
measured in such a complicated environment.  This result will be
submitted shortly to Phys. Rev. D Rapid Communications and we
anticipate updates to the measurement that make use of more data in
the near future.





\begin{thebibliography}{9}
\bibitem{Ref:CC}
Charge-conjugate states are implied throughout the paper unless
otherwise specified.
\bibitem{Ref:TTT} 

Two tracks of opposite charge are required to have an impact parameter
$0.012 \le d_0 \le 0.1$~cm, individual transverse momentum $p_T >
2.0~\gevc$, $p_{T,1} + p_{T,2} > 5.5~\gevc$, and an opening angle
$\Delta\phi < 135^{\circ}$.
\end{thebibliography}


\end{document}